\documentclass[12pt,preprint]{aastex}
\slugcomment{version: noon, 15 January 2003}
\shortauthors{Fabbiano et al.}
\shorttitle{IC~1459 nuclues with {\it Chandra}}

\def\ergcm2s{erg cm$^{-2}$ s$^{-1}$} 
\def\etal{et al.}		
\def\chandra{{\it Chandra}}

\begin{document}

\title{The X-ray-faint Emission of the Supermassive Nuclear Black
Hole of IC~1459}

\author{G. Fabbiano$^1$, M. Elvis$^1$, S. Markoff$^2$,
A. Siemiginowska$^1$, S. Pellegrini$^3$, A. Zezas$^1$,
F. Nicastro$^1$, G. Trinchieri$^4$, J. McDowell$^1$}

\bigskip

\affil{1. Harvard-Smithsonian Center for Astrophysics, 60
Garden Street, Cambridge, MA 02138}
\email{pepi@cfa.harvard.edu, elvis@cfa.harvard.edu,
aneta@cfa.harvard.edu, azezas@cfa.harvard.edu,
nicastro@cfa.harvard.edu, jcm@cfa.harvard.edu}
\affil{2. Center for Space Research, MIT, Cambridge, MA 02139}
\email{sera@alum.mit.edu}
\affil{3. Universit\'{a} di Bologna, Italy}
\email{pellegrini@bo.astro.it}
\affil{4. Osservatorio di Brera, Milano, Italy}
\email{ginevra@brera.mi.astro.it}

\bigskip

\bigskip
\begin{abstract}

\chandra\ observations of the supermassive black hole in the
nucleus of IC~1459 show a weak
($L_X$=8$\times$10$^{40}$erg~s$^{-1}$, 0.3-8~keV), unabsorbed
nuclear X-ray source, with a slope $\Gamma$ = 1.88$\pm$0.09, and
no strong Fe-K line at 6.4~keV (EW$<$382~eV). This describes a
normal AGN X-ray spectrum, but lies at 3$\times$10$^{-7}$ below
the Eddington limit. The SED of the IC~1459 nucleus is extremely
radio loud compared to normal radio-loud quasars.
The nucleus is surrounded by hot ISM (kT$\sim$0.5-0.6~keV) with an
average density of 0.3~cm$^{-3}$, within the central $\sim$180~pc
radius, which is comparable to the gravitational capture radius,
$r_A\sim$140~pc. We estimate that for a standard AGN efficiency of
10\%, the Bondi accretion would correspond to a luminosity of
$\sim$6$\times$10$^{44}$erg~s$^{-1}$, nearly four orders of magnitude
higher than $L_X$.
ADAF solutions can explain the X-ray spectrum, but not the high
radio/X-ray ratio. A jet model fits the radio-100$\mu$m and
X-ray spectra well. The total power in this jet is $\sim$10\% of
L$_{Bondi}$, implying that accretion close to the Bondi rate is
needed.
%

%

\end{abstract}


 \clearpage
\section{Introduction}

Studies of the stellar dynamics of the inner cores of galaxies
have established the presence of hidden supermassive objects of
$10^7 - 10^9 M_{\odot}$, which are believed to be black holes
(Richstone et al 1998, Magorrian et al 1998, van der Marel 1999).
The masses of these black holes are loosely correlated with the
bulge luminosity (Magorrian et al 1998) and tightly correlated
with the central velocity dispersion (Ferrarese \& Merrit 2000,
Gebhardt et al. 2000) and the central light concentration (Graham
et al 2001), and appear to be ubiquitous in giant E and S0
galaxies. Since accreting black holes are believed to be
responsible for luminous active galactic nuclei (AGN) and
quasars, the question arises of why nuclear activity is not more
widespread, instead of being confined to only a small percentage
of all galaxies.  Suggested explanations for this lack of nuclear
activity have included the presence of Compton-thick obscuring
material in the line of sight to the nucleus (e.g. the Circinus
galaxy, Matt et al. 1999), feedback modulated accretion with a
low duty cycle (Binney \& Tabor 1995, Ciotti \& Ostriker 2001),
and inefficient accretion onto the black hole (e.g., Sgr A*, see
review by Melia \& Falcke 2001, Di~Matteo et al. 2000). The
latter include instability-driven periodic accretion in a thin
accretion disk (Siemiginowska \& Elvis 1997; Siemiginowska,
Czerny \& Kostyunin 1996), and the more popular
radiatively-inefficient advection dominated accretion flow (ADAF,
CDAF, ADIOS) models (see review by Narayan 2002).

X-ray observations offer a way of constraining these models, and
discriminating among different emission scenarios, because
different characteristic X-ray spectra are expected in each case.
However, until recently, it was impossible to get a `clean' look
at the faint AGN that may be hidden in the nuclei of large bulge
galaxies, because of the relatively large beams of X-ray
telescopes ($> 5''$~FWHM).  \chandra\ (Weisskopf et al. 2000),
with its $\sim 0.3"$~FWHM PSF (Van Speybroeck et al. 1997), gives
us the unique opportunity to take a direct, uncontaminated, look
at the faint nuclear emission that may be associated with
supermassive nuclear black holes. Moreover, with \chandra\ we can
investigate directly the circum-nuclear region down to $\sim
100$~pc radii, or even deeper in, depending on the distance of
the galaxy. This gives us a direct way to constrain the accretion
rate on the black hole, if a hot interstellar medium (ISM) is
detected (Bondi 1952).

In this paper we report the results of the \chandra\ ACIS-S
observation of the nucleus of IC~1459 (Table~1).  IC~1459 is a
much studied Elliptical (E3) galaxy. This galaxy has a
counter-rotating core (Franx \& Illingworth 1988), suggesting a
merger event in its past, and a massive nuclear black hole ($2
\times 10^9 M_{\odot}$, see Table~1), which is however associated
only with moderate AGN activity.  The nucleus of IC~1459 has a
strong (1~Jy) largely compact radio source with diameter
$<$0.03arcsec ($<$3~pc for a distance D = 22~Mpc, Slee et al.,
1994, Ekers et al. 1989) and an inverted spectrum ($\alpha = -
0.21$, $F_\nu \propto \nu^{-\alpha}$, Drinkwater et al
1997); a hard AGN-like X-ray component (ASCA; Matsumoto et al
1997); a strong LINER optical spectrum (Phillips et al., 1986);
and the large 25$\mu$m/60$\mu$m IRAS ratio typical of AGN
(de~Grijp et al., 1985).  However, a normal radio-loud AGN with
this core 5~GHz flux (Elvis et al., 1994) would have an X-ray
luminosity some 10~times brighter than the 10$^{41}$erg~s$^{-1}$
observed in IC~1459 (Fabbiano, Kim \& Trinchieri 1992), and a
non-thermal optical continuum at least 2~magnitudes brighter than
observed. Instead this E3 galaxy has photometrically normal
colors down to a few arcsec and a normal elliptical surface
brightness profile into the central 1~arcsec ($\sim$107~pc for
the distance in Table~1, Franx et al., 1988).

To investigate the X-ray properties of this nucleus, we observed
IC~1459 with \chandra\ ACIS-S. This observation allows us to separate
the nuclear point source from the surrounding galaxian emission, and
thus obtain an uncontaminated characterization of its X-ray
spectrum. We can also explore the properties of the circum-nuclear
region, that may provide the fuel for the AGN.  Here we report the
results of this observation as it pertains to the nuclear source. The
X-ray properties of the general galaxian emission will be the subject
of a future paper.  In \S2. we describe the data reduction and the
spectral analysis; in \S3. we compare these results with model
prediction.

\section{Observations and Data Analysis}

\chandra\ was pointed to IC~1459 on 12 August 2001, for 58.8~ks
with the back-illuminated ACIS-S3 CCD chip at the focus
(Observation ID: 2196).  Table~1 summarizes some of the
properties of IC~1459 and gives the log of the \chandra\
observations.  The observations were done in a 1/2 sub-array
mode, to minimize possible `pile-up' (\chandra\ {\em Proposers'
Observatory Guide} 2001) of a bright nuclear source. This
resulted in halving the effective chip area, but given the galaxy
size (5.2$\times$3.8~arcmin, NED), the entire optical extent of
IC~1459 is still imaged.  ACIS was at a temperature of -120 C
during these observations.  The satellite telemetry was processed
at the \chandra\ X-ray Center (CXC) with the version 6.2.2 of the
Standard Data Processing (SDP) system and CALDB version 2.6, to
correct for the motion of the satellite and to apply instrument
calibration.  Verification of the data products showed no
anomalies.  The data products were then analyzed with the CXC
CIAO software (Version 2.2.1). CIAO Data Model tools were used
for data manipulation, such as screening out high background
data, and producing images in given energy bands. DS9 and
`funtools' were used for interactive image analysis.

The data were screened to exclude high background regions,
following the thread in the CIAO web page 
\footnote{\tt http://cxc.harvard.edu/ciao/threads/filter\_ltcrv/}
which results in the removal of data with global background
values outside $\pm 3 \sigma$ of the mean.  Because this
procedure resulted in 3 Good Time Interval (GTI) files, these
files were merged with {\sc dmmerge} and then applied together
with the original GTI from processing. The resulting screened
exposure is 55.1~ks. Since a new ACIS-S3 calibration was released
after our initial analysis, we have applied it to the event file,
by running `acis\_process\_events'. Using this recalibrated data,
we found that any differences in our spectral results are
negligible (at the second decimal place) and well within the
errors.

Fig.~1 (left) shows the (0.3-10~keV) ACIS-S3 image of the central
area of IC~1459. A point-like source is prominent in the nucleus.
A number of fainter point-like sources, in all likelihood
belonging to the XRB population of IC~1459, are also visible, at
significantly lower count rates; these will be the subject of
future work. Fig.~1 (right) shows an adaptively smoothed image of
the same region (using {\sc csmooth}). This image shows clearly the
diffuse emission in the field, including a luminous flattened E-W
elongated region surrounding the nucleus.  We will discuss the
extended emission in detail in the future. Here we concentrate on
the properties of the point-like nuclear source.

\subsection{The Nuclear Spectrum}

We used the `acisspec' script, in the 2.2.1 CIAO release, to
extract the nuclear data for spectral analysis. This task
produces spectral response matrices files (RMF), by a weighted
average (based on the counts for each ACIS pixel) of all the
relevant calibration files. The background was taken from a large
source-free region in the portion of the ACIS S3 chip unoccupied
by IC~1459 (a circle of 78" radius, centered 205" SW of the
nucleus). This background is representative of the field
background, and does not include the contribution of the local
galaxian diffuse emission (see fig.~1) in the nuclear beam. The
effect of vignetting at these radial distances is less than 10\%.
We chose this approach, rather than subtracting a local
background, because the surface brightness of the diffuse
galaxian emission is likely to be larger at the inner radii than
in the surrounding regions.  So, considering this emission
component explicitly in the spectral fit of the nuclear data is
in principle less conducive to error.  The nuclear data were
extracted from a 1.7" radius circle centered on the centroid of
the nuclear count distribution [RA = 22 57 10.62, Dec = -36 27
43.9, (J2000)], shown in fig.~1 (left panel). This radius
contains $\sim 90\%$ of the nuclear PSF ({\em Chandra Proposers'
Observatory Guide}, 2001).  The data were binned in energy bins
with a minimum of 15 net counts per bin.  We restricted our
spectral analysis to data in the 0.3-8~keV range, because the
spectral calibrations are more uncertain at the lowest energies,
and the instrumental background tends to dominate at the highest
energies.  Even so, the background contributes less than a count
in our high energy spectral bins. We obtained a total of 6544.7
net counts distributed into 198 spectral bins.  We used {\em Sherpa}
to fit the spectral data to models. The results are summarized in
Table~2.

We first fitted the data with an absorbed power-law
model. Although this fit was good (see Table~2), we noticed a
possible trend of excess residuals at the energies $\sim 0.5 -
1$~keV. This can be seen from fig.~2a, which shows the
0.3-2.0~keV portion of the spectrum, with the best fit and
residuals.  Restricting the fit to the 0.3-1.3~keV range, we find
that the simple absorbed power law model is not a good
representation of the data in this energy range ($\chi^2 = 78.4$
for 59 degrees of freedom).  To account for these deviations, we
used a composite model consisting of an absorbed power-law plus
optically thin plasma emission (MEKAL, the Mewe-Kaastra-Liedhal
plasma model; e.g. Liedhal et al 1995) with solar abundance, and
applied (fixed) Galactic $N_H$ to both components. This choice is
justified by the presence of diffuse emission in the
circum-nuclear region (fig.~1), which would contribute to the
spectral counts, given that the background was derived from a
region outside the galaxy.  This model was first fitted to the
entire 0.3-8~keV spectrum; the resulting $\chi^2$ (Table~2) is
155 for 192 degrees of freedom, suggesting that the model may
`over-represent' the data. However, when the best fit model is
applied to the 0.3-1.3~keV data, to estimate the goodness of fit
in this spectral range, an F test shows that this is now improved
at the 99.5\% confidence level (almost $3 \sigma$).  In this
composite model, a thermal component with kT$\sim 0.6$~keV and
X-ray luminosity of $\sim 2 \times 10^{39} \rm ergs~s^{-1}$ (D =
22~Mpc) accounts for the soft excess. The power-law slope
($\Gamma \sim 1.9$) and unabsorbed luminosity ($\sim 8 \times
10^{40} \rm ergs~s^{-1}$) do not change appreciably
(Table~2). Fig.~2b shows the data, best-fit 2-component model and
residuals in the 03.-2.0~keV range.  Fig.~3 shows the data, fit
and residuals for the 2-component model for the 2.0-8.0~keV
portion of the spectrum. Fig.~4 shows the power-law component
$\Gamma$ and $N_H$ confidence contours.  Fig.~5 shows
kT-normalization confidence contours for the thermal component.

Visual inspection of the spectrum (fig.~3) does not reveal any
excess emission at 6.4~keV, to suggest the presence of an Fe
K-$\alpha$ line. We derived a 3$\sigma$ upper limit of $<$382~eV
on the equivalent width of such a line over the best-fit
power-law, by forcing the fit of a narrow line at 6.4~keV,
calculating the 3$\sigma$ upper bound, and then using the
{\em Sherpa} {\sc eqwidth} command to calculate a 3$\sigma$
equivalent width over the best-fit power-law continuum.

\subsection{The Circum-Nuclear Hot ISM}

As shown by fig.~1, the circum-nuclear environment in IC1459 is
pervaded by extended emission, most intense in the $\sim 2''$
annulus surrounding the nuclear spectrum extraction region.  The
nuclear spectral fit above suggests the presence of hot ISM with
kT$\sim$0.6~keV. Such a hot ISM may fuel the central black hole.
Following Bondi accretion theory (1952), we can estimate the
accretion rate at the gravitational capture radius, which depends
on the mass of the black hole and the temperature and density of
the ISM (see e.g. Di~Matteo et al 2002).  The ISM parameters can
be estimated from the X-ray data.

To constrain the temperature and density, we investigated the
spectral properties of the hot ISM, by analyzing the spectrum of
data extracted from a point-source-free circum-nuclear polygon
(fig.~1, right panel), in the 0.3-8.~keV range.  The results are
summarized in Table~3. A two-component model (optically thin
thermal emission and a power-law) was required to obtain an
acceptable fit (see also discussion in Kim \& Fabbiano, 2002, for
a similar situation in NGC~1316). The power-law normalization is
$3\%$ of that of the nuclear power-law (equivalent to $\sim 2.5
\times 10^{39} \rm ~ergs~s^{-1}$), consistent with spill-over of
nuclear counts.  However, unresolved galaxian X-ray binaries
could also contribute (see Kim \& Fabbiano 2002); these issues
will be examined fully in a future paper.  Of immediate
relevance, the best-fit temperature of the thermal component is
consistent with the results of the two-component nuclear fit,
confirming the presence of a hot ISM with kT$\sim 0.5 - 0.6$~keV
in the vicinities of the nuclear black hole.

Because of the overpowering presence of the nuclear point-source,
we cannot directly measure the luminosity of the hot ISM in the
centermost region, from which to obtain an estimate of the
electron density $n_e$. We have an estimate of this luminosity
from our spectral analysis (Table~2, \S2.1), but it includes a
foreground and background contribution from the surrounding hot
ISM.  Since {\it Chandra} resolves this emission, we can estimate
its contribution and subtract it. To this end, we used an
extraction region comparable in size and adjacent to the nuclear
one to estimate the amount of emission from the surrounding
volume. We subtracted this value from the nuclear thermal
emission estimate, to obtain the emission from the circum-nuclear
volume. This procedure results in a corrected luminosity of $1.6
\times 10^{39} \rm ergs~s^{-1}$ from a sphere corresponding to
the 182~pc radius nuclear spectrum extraction region, which is
comparable to the 140~pc gravitational capture radius (\S 4.1).
This luminosity is 20\% lower than the hot ISM $L_X$ in
Table~2. A minimum lower estimate on this value could be obtained
by a simple extrapolation of the background-subtracted value from
the circum-nuclear region. This would amount to $2.6 \times
10^{38}~ \rm ergs~s^{-1}$. We used these estimates to correct the
emission measure from the spectral fit (Table~2) to derive $n_e =
0.31~\rm cm^{-3}$ ($n_e = 0.12~ \rm cm^{-3}$, if the minimum
lower estimate of $L_X$ is used).  We estimate the accretion rate
based on the above parameters and discuss its consequences below.

\section{Spectral Energy Distribution of IC~1459}
\label{sed}

We have compiled literature data to complement our results and
derive the spectral energy distribution (SED) of IC~1459
(Table~4, fig.~6).
The radio source (Slee et al. 1994) has a size $<$0.03~arcsec,
and the X-ray nuclear source is compact at the arcsec level, so
their total fluxes represent the nucleus well.  In the millimeter
band the size of the source is unknown ($<$17-28 arcsec, Knapp \&
Patten 1991), but the total fluxes appear to form an
extrapolation of the radio spectrum, so that the source is likely
to be compact.
In the far-infrared the large beam (arcminute) 12$\mu$m-100$\mu$m
IRAS fluxes lie above the radio-mm spectrum. A small beam
(5~arcsec dia.) 10$\mu$m point (Sparks et al. 1986) has only 10\%
of the IRAS 12$\mu$m flux. So although IC~1459 is unusual in
having detections in all four IRAS bands, it appears that most of
this emission originates outside the nuclear region, at least at
the shorter wavelengths.

There could be cooler dust to which IRAS was not sensitive.  The
most possible dust would be given by fitting the $\nu^4$ tail of
a dust spectrum (in $\nu f_{\nu}$ space) through the highest
frequency millimeter photometry point at 0.8~mm.  In this case
this fails to fit the lower frequency millimeter points at 1.1,
1.3 and 2.0~mm.  So up to 2.0~mm non-thermal processes must be
dominant, but the higher frequency infrared points can be
ascribed to dust in the galaxy body.
The small aperture (5 arcsec dia.) JHK measurements lie well
above the equal aperture 10$\mu$m point.  

A blue (V-I$\sim$1, Carollo et al. 1997) point-like nucleus with
V=18.4 is seen in HST imaging (Tomita et al. 2000, Cappellari et
al. 2002). This color is the same as that of 3C~273 (McAlary et
al. 1983).  The blue nucleus is close to a narrow band of
`strong' dust absorption. An F814W band HST image of the IC~1459
nucleus is giving a nuclear flux of 0.26~mJy (Verdoes-Kleijn et
al. 2002).

For comparison we plot in fig.6 the median SED of the radio loud
low redshift quasars from the sample of Elvis et al. (1994). This
SED has been renormalized arbitrarily to match the nuclear
optical data for IC~1459, comes close to the X-ray flux, but
falls well below the radio data. The IC~1459 nucleus is thus
extremely radio-loud compared even with radio-loud quasars.

IC~1459 has both nuclear X-rays and nuclear H$\alpha$ emission
(Verdoes-Kleijn et al. 2002) at a level
$log~L(H\alpha)=$37.9~erg~s$^{-1}$ (Macchetto et al. 1996, corrected
from their distance of 29~Mpc to our value of 22~Mpc).  These values
fit on the normal $L_X$~vs.~$L(H\alpha)$ correlation (Elvis et
al. 1978, Ho et al. 2001).  Using $L_{H\alpha}$ to predict the
ionizing continuum (Osterbrock 1989, equation~11.3, figure~11.6), and
assuming a power-law continuum slope of 1.5, gives
$f$(912\AA)=0.021$f_C^{-1}$~mJy, where $f_C^{-1}$ is the covering
factor of continuum source for the H$\alpha$ emitting gas. The derived
$f$(912\AA) would be a factor 2 larger for a continuum slope of 2.0.

\section{Discussion}

\subsection{Luminosity and Bondi Accretion}

The X-ray luminosity of the nuclear point-source in IC1459
(7.9$\times 10^{40} \rm ~ergs~s^{-1}$) is $\sim 3 \times 10^{-7}$
of the Eddington luminosity of the 2$\times 10^9 M_{\odot}$
nuclear black hole (Cappellari et al. 2002, for D = 22~Mpc).  Why
is this nucleus not a luminous AGN? In principle, since the AGN
is surrounded by a hot ISM, there should be enough fuel to power
the black hole and generate higher X-ray luminosity.  However,
this requires (1) the gas to accrete onto the black hole and (2)
relatively high efficiency of the accretion process.

Following Fabian \& Canizares (1988) and Di~Matteo et al. (2002),
we can estimate the Bondi accretion rate using the parameters of
the hot ISM derived in \S2.  Bondi theory is used to obtain an
estimate of the gravitational capture radius (see also Frank,
King \& Raine 1992): $r_A~=~0.05~T_{0.8}^{-1}~M_9~\rm{kpc}$,
where $T_{0.8}$ is the temperature of the ISM in units of
0.8~keV, and $M_9$ is the mass of the black hole in units of
$10^9 M_{\odot}$.

For the parameters of the nucleus of IC~1459, the gravitational
capture radius is at $r_A \sim$140~pc. While at the distance of
IC~1459 the combined \chandra\ mirror and ACIS resolution (0".5)
corresponds to 53~pc, and so in principle a direct measurement of
the gas parameters at the capture radius could be attempted, this
is impeded by the presence of the bright nuclear point-like
source, which dominates the emission within a
1$^{\prime\prime}$.7 radius (\S2).  The density at the capture
radius is likely to be higher than our average estimate from
\S2.2, given that the radial density distribution tends to
increase at smaller radii (see e.g. M87, di Matteo et al 2002).
Therefore our estimate of the Bondi accretion rate ($\dot
M_{Bondi}$) can be considered as a lower limit.

We estimate $\dot M_{Bondi}$ from the expression given in Di
Matteo et al (2002):
\newline
$7\times 10^{23}~M_9^2~T^{-3/2}_{0.8}~n_{0.17}~\rm g~s^{-1}$,
where $n_{0.17}$ is the density of the ISM in units of
0.17~cm$^{-3}$.  We adopt kT=0.5~keV, as suggested by our
spectral analysis, for the temperature of the ISM, and $n_e =
0.31~ \rm cm^{-3}$, as derived in \S2.2, and obtain $\dot
M_{Bondi} \sim 0.16~ M_{\odot}$year$^{-1}$.  Using our extreme
lower estimate of $n_e = 0.12~ \rm cm^{-3}$ (see \S2.2), would
result in reducing this accretion rate by a factor of 2.5 to
about 0.06 M$_{\odot}$year$^{-1}$. Plausible variations in kT
would only have a minor effect; Chandra observations do not find
dramatic cooling in galaxy cores (e.g. Kim \& Fabbiano 2002,
David et al. 2001, Kaastra et al. 2001).

The above accretion rate is similar to the accretion rates
required to power high luminosity quasars.  The Eddington
luminosity for the black hole mass in IC1459 is of order
2.5$\times 10^{47}$~ergs~s$^{-1}$, while the observed luminosity
is 3$\times 10^{-7}$ times lower. If the efficiency of the
accretion flow was $\sim 10\%$, as generally assumed in the
standard accretion disk theory, then the luminosity of IC1459
would be in the range $10^{44}-10^{45}$~ergs~s$^{-1}$, a normal
AGN luminosity. So why is the observed luminosity so low?

Assuming steady accretion, the radiative efficiency, $\eta =
{{L_{accr}} \over {\dot M_{Bondi} c^2}}$ where $L_{accr}$ is the
observed luminosity of the nucleus. For L$_{accr}$=7.9$\times
10^{40}$~ergs~s$^{-1}$ and $\dot
M_{Bondi}$=0.16~M$_{\odot}$~year$^{-1}$ we obtain $\eta=8.5
\times 10^{-6}$. This efficiency is much lower than in the
standard accretion onto black hole scenario.  In the following
sections we discuss possible ways of explaining the observed
luminosity:
(1) hiding the emission, with gas and dust, 
(2) impeding the accretion into the black hole,
(3) using inherently low radiative efficiency processes, 
(4) removing energy from the power available for radiation
losses, e.g. in a jet.
%

\subsection{Hiding the Emission: Obscuration}

Does the `missing' luminosity come out at different wavelengths?
We can search for obscuration effects both from the X-ray
spectrum, and by considering the complete SED of IC~1459.

There is no strong photoelectric absorption signature at low
energies in the X-ray spectrum. To suppress the X-ray spectrum by
a factor 100 requires a column density, N$_H>$10$^{23}$cm$^{-2}$,
and such values are seen in AGN known to be heavily obscured
(e.g. NGC~1068, Ueno et al. 1994; Circinus, Matt et al. 1999).  The
absence of strong low energy absorption however is not sufficient
to conclude that obscuration is not at work.  In some cases the
direct spectrum is completely obscured, and only a small
scattered fraction can be seen (NGC~1068, Circinus). In all these
cases though, the scattered spectrum contains strong fluorescent
emission lines, notably of Fe-K at 6.4~keV, with EW$>$1~keV. This
is excluded by the 382~eV 3$\sigma$ upper limit on Fe-K in the
{\em Chandra} spectrum.

Obscuration would be in agreement with our X-ray results, if the
nucleus were embedded in a thick spherical dust distribution, that
would let escape only a very small amount of the emitted power.  An
irregular dust distribution is present in the nuclear region of
IC~1459 (Goudfrooij et al., 1990). Although on average this provides
too little reddening [E(B-V)=0.07, or 4$\times$10$^{20}$cm$^{-2}$ for
standard dust to gas ratio and composition] to hide an AGN, the
obscuration could well be much greater near the nucleus (Tomita et
al. 2000).
In this case, however, we would expect to see the `missing'
luminosity in the infrared.
Within a 5~arcsecond aperture the SED distribution in the
1-10$\mu$m IR (figure~6) follows a Rayleigh-Jeans tail quite
closely and is likely to be stellar. If the large beam IRAS 
10-100$\mu$m FIR emission were nuclear dust emission, 
as argued by Walsh et al. (1990) on the basis of a lack of
21~cm HI line emission, this would not account for the sub-Bondi
X-ray luminosity, since the total FIR luminosity is $\sim$60
times the observed X-ray luminosity, not the factor 7500
required.

\subsection{Impeding Accretion}

Bondi theory applies to spherical accretion. However, the
accreting gas can have a significant amount of angular momentum.
The gravitational capture radius for IC1459 is at 140~pc =
7.4$\times 10^5 r_g$.  A standard disk is gravitationally
unstable at large radii ($>$ a few 1000$r_g$). So there is a
factor $>$100 difference in radius between where we have measured
\.{M} to where any disk could be.  Formation of an accretion
disk from a spherical flow is not well understood as yet, but
depends on the angular momentum and the temperature of the gas at
the gravitational capture radius.  
Igumenshchev, Illarionov \& Abramowicz (1999) simulated an accretion
flow with low angular momentum accreting matter and show that the
matter condenses into a cool thin disk. So a two component (hot and
cold) plasma can exist in the vicinity of the black hole. The size of
the cold thin disk is smaller at higher accretion rates. The hot
plasma surrounds the disk and extends up to the outer gravitational
capture radius.  The accretion rates in the cold and hot phase are
likely to be different, although no calculations exist.
Hawley and Balbus (2002) provide a dynamical picture of the 3D
non-radiative accretion flow, where the flow has three well
defined components: a hot, thick Keplerian disk, surrounding
a magnetized corona with circulation and outflow, and a
magnetically confined jet. The accretion disk is very hot and
forms a toroidal structure in the innermost (r$< 10r_g$)
regions. They estimated the energy output for this model and
found consistency in the case of SgrA* data.  However, their
simulations do not include any radiation processes and are only
run within the central few hundred $r_g$, well inside the Bondi
radius.

The accretion flow geometry depends on the physical conditions of
the matter in the region where the gas starts to be influenced by
the black hole.  Abramowicz \& Zurek (1981) discussed the
geometry of the adiabatic accretion flow and show that for
sufficiently high angular momentum the flow forms a disk-like
pattern, while quasi-spherical transonic accretion flow is
possible for the matter with small angular momentum. Similar
results of bimodal geometry were recently discussed by Yuan
(1999), who shows that the outer boundary conditions (e.g. gas
temperature, velocity, angular momentum) are critical to the type
of the accretion pattern in the viscous, optically thin flow.
The accretion rate could be then significantly less than the
Bondi rate if the accreting gas has a significant amount of
angular momentum.  This could result in inefficient accretion at
the Bondi gravitational capture radius $r_A$. Even if Bondi
accretion is effective, angular momentum could stop the flow
closer in, if a disk forms and angular momentum cannot be
transferred effectively outwards.  Removing angular momentum from
gas at $\sim$10~pc is a longstanding problem in fueling AGN
(Blandford 1990), so stalling accretion at or near the Bondi
radius is not hard.

Assuming that the gas can reach accretion disk-like radii
($<$10$^4~r_s$, $\sim$1~pc, where $r_s$ is the Schwarzschild
radius, for $M_{\bullet}=2\times 10^9~M_{\odot}$), there are
several ways to inhibit accretion.  Two types of accretion disk
could form near the black hole: standard optically-thick
geometrically-thin disks, or optically-thin thick disks
(advection dominated accretion flows, so-called ADAF).  Standard
thin accretion disks can be thermally and viscously unstable and
undergo irregular outbursts (Lin \& Shields 1986, Mineshige \&
Shields 1990, Siemiginowska, Czerny \& Kostyunin 1996) leading to
quiescent periods with low accretion efficiency.  In the ADAF
family (discussed in \S\ref{adaf}), convection dominated
accretion flows (CDAFs, Quataert, E., \& Gruzinov, 2000, and
reference therein) also stall accretion at small radii by
convecting the material back out to larger radii. Again this is a
temporary effect, although how long a CDAF can prevent accretion
has not yet been studied.  Advection dominated inflow/outflows
solutions (ADIOS, Blandford \& Begelman 1999) also prevent
accretion in the inner regions, but do so by removing matter from
the inflow completely via a polar outflow. Thus an ADIOS may be
one mechanism of pairing a radiatively inefficient accretion flow
to the origin of AGN jets (see Yuan, Markoff \& Falcke 2002 for a
discussion relevant to Sgr A*).

\subsection{Low Radiative Efficiency Models}
\label{adaf}

Radiatively inefficient scenarios include `advection dominated
accretion flows' (ADAFs), `convection dominated accretion flows'
(CDAFs), or `advection dominated inflow outflow solutions'
(ADIOS) (see review by Narayan 2002), in all of which the matter
in the center becomes so hot and tenuous that it is unable to
radiate strongly. In pure inflow ADAFs much of the energy is
carried by the less radiative protons and advected inside the
event horizon. In ADIOS (Blandford \& Begelman, 1999) the matter
flows outward before cooling (possibly radiating, via magnetic
fields, as a jet).  In CDAFs (Begelman \& Meier 1982) the matter
convects back out to larger radii. Variations of these
models have had success accounting for the spectra of
low-luminosity AGN, such as the quiescent state of the nucleus of
our spiral galaxy, Sgr A* (see review of models in Melia \&
Falcke 2001).

ADAFs have also been invoked to explain the SED of the nuclei of
similar elliptical galaxies, which have hard X-ray sources
(L$_X\sim$10$^{40}$erg~s$^{-1}$) strong inverted spectrum radio
sources, but no non-stellar optical continuum (e.g. NGC~1399,
M~87, Allen, Di~Matteo \& Fabian, 2000; Di~Matteo et al. 2000).
However these consistently overpredict the observed radio
emission in normal, weak radio, ellipticals (Lowenstein et
al. 2001, Di~Matteo, Carilli \& Fabian 2001).

In IC~1459 we can exclude that the X-ray emission is dominated by
a pure bremsstrahlung in a pure inflow, low accretion rate, ADAF
(Narayan \& Yi 1994, 1995), since that would have a flat X-ray
power-law ($\Gamma = 1.4$) and no Fe-K line.  While no Fe-K line
is seen in the {\em Chandra} spectrum, the slope is steeper than
expected ($\Gamma$=1.88$\pm$0.09, Table~2). The steep power-law
is consistent with the prediction of the ADAF model for higher
accretion rates, comparable to those we estimate for IC~1459.  In
this case, an X-ray component from Comptonized radio synchrotron
emission is generated (Esin et al. 1998), as for example has been
suggested in the case of M87 (Di~Matteo et al 2001).  Also,
convection dominated accretion flows (CDAFs) give $\Gamma
\sim$1.6-2.0 for the L$_X$/L$_{Edd} \sim$10$^{-7}$ value we see
in IC~1459 (Ball, Narayan \& Quataert 2001).

However, following the argument by Pellegrini et al (2003) for
IC~4296, a pure inflow ADAF model cannot explain the entire
radio-to-X-ray nuclear emission of IC~1459. M~87 and IC~1459 have
comparable M$_{\bullet}$ and \.{M}, the two primary ADAF variables, so
a comparison of their SEDs is instructive.  The ratio of X-ray to
radio emission is well constrained in these models ($\nu
L_{\nu}$(1~keV)$/\nu L_{\nu}$(22~GHz)$ \sim$60 in M87 (Di~Matteo et al
2002). In IC~1459 the ratio is significantly smaller, $\nu
L_{\nu}$(1~keV)$/\nu L_{\nu}$(22~GHz)$\sim$3 (fig.~6), so the ADAF
predicted radio emission would be much less than observed.  Even
larger X-ray/radio ratios would be expected for CDAF and ADIOS models
(Quataert and Narayan 1999).  This does not exclude the possibility of
an ADAF or CDAF as the origin of the X-ray emission, but in all these
cases a different source, for example jets, would be required to
explain the radio. This would be consistent with the surveys of
low-luminosity AGN, which suggest that the jet is dominating at least
the radio emission in these systems (Nagar, Wilson \& Falcke 2001).

\subsection{A Jet-Disk Model for the non-thermal X-ray/radio
Continuum} 
\label{markoff}

We have fitted the jet model of Markoff et al. (2001, 2003) to
the SED for IC~1459.  The details of the model can be found in
these papers, and references therein, so we give only a brief
summary below.

The basic idea of this model is that a fraction of neutral
electron-proton plasma from the accretion flow is advected into a
jet, which is freely expanding and accelerated via its
longitudinal pressure gradient.  Eventually, the plasma
encounters an acceleration region (assumed for simplicity to be a
shock) and some of the thermal plasma is redistributed into a
power-law distribution.  The main fitted parameters are the power
in the jet, characterized as a fraction of the Eddington power,
the location where the acceleration zone begins, the energy index
of the accelerated electrons and the inclination angle.  A
multi-color blackbody thin accretion disk (Shakura \& Sunyaev
1973; Mitsuda et al. 1984) is assumed to contribute both to the
spectrum directly, as well provide a photon field for inverse
Compton upscattering by the jet plasma.  This disk is
parameterized by its inner temperature and its thermal luminosity
in Eddington units.  The thermal disk spectrum as well as
synchrotron, synchrotron self-Compton (SSC) and external (disk)
Compton (EC) emission from the jet are all calculated.

The resulting fit is shown in fig.~7, and the fitted model
parameters are given in Table~5. The post-shock synchrotron
spectrum fit to the radio continuum is good to factors of
$\sim$2. The extrapolation of the optically thin part of this
post-shock synchrotron spectrum fits both the X-ray slope and
normalization notably well. The pre-shock synchrotron emission
fits the 2~mm to 60~$\mu$m peak, with the shorter wavelength IR
emission being assumed to come from dust in the body of the
galaxy, as evidenced by the large fraction of the 10~$\mu$m
emission coming from outside a 5~arcsec aperture. The HST optical
photometry of the nucleus is fitted by a cool accretion disk.

The two bumps in the Compton component (see fig. 7) come from SSC
within the nozzle plus EC from the weak disk blackbody.  The
synchrotron power law comes from much further out, 700~$r_g$, and
so is not as strongly upscattered.  The joint SSC/EC component
cuts out at low frequencies because, for computational
efficiency, this component is only calculated out to where it
becomes too low to matter, a few 100 times the nozzle length.

Interestingly, the location of the shock region falls exactly in
the range that Markoff, Falcke \& Fender (2001) and Markoff et
al. (2003) have been finding for X-ray binaries ($z_{sh}\sim
10-1000 r_g$), and is also consistent with values found for BL
Lacs (e.g. Beckmann et al. 2002).  This suggests that there may
be commonality in the location of the acceleration region even
for jets of vastly different scales.

There may well be other sources of EC emission, e.g. another source of
external photons to be Compton scattered, such as the broad emission
lines, or a hotter disk (see below). The X-rays may well be then a
varying mix of synchrotron and EC at different energies, with
synchrotron being more important at low energies.

While the blackbody disk component produces only about
10$^{-7}~L_{Edd}$ in luminosity, the jet carries
10$^{-4}~L_{Edd}$ in kinetic and internal particle and magnetic
field energy in this model. I.e. the jet is by far the dominant
sink of power, although only a small fraction ($\sim$3\%,
7.6$\times$10$^{-6}~L_{Edd}$) is radiated away.  A small caveat
is that, since the X-ray slope is rising in $\nu f\nu$, the X-ray
power radiated depends on the maximum cutoff for the synchrotron
emission. But this is unlikely to increase the radiated fraction
substantially.  Also, the total jet power is somewhat dependent
on the inclination angle assumed, because of beaming. If the
inclination angle were closer to 60 degrees, then the total jet
power required to fit the spectrum would be a larger fraction,
but never more than $\sim$20-30\%. For this fit the total jet
power is then 10\% of the Bondi rate (2.25$\times$10$^{-3}$
L$_{Edd}$), i.e.  $q_{jet}$=0.1. This is a much larger fraction
of L$_{Edd}$ than had been thought from estimates based only on
radiative power, but is in line with the $q_{jet}$ values of
0.1~-~0.001 found for models of Galactic binaries (Markoff et
al. 2001, 2003).  This suggests that the Bondi rate is a more
realistic estimate of the accretion rate than had been claimed.

This brings up two interesting points.  First, these results seem
to support the recent findings of, e.g., Nagar, Wilson \& Falcke
(2001), whose surveys are discovering that jets seem capable of
dominating the disk emission in low-luminosity systems.  This is
also seen in the high ratio of jet to radiated power required in
the Yuan, Markoff \& Falcke (2002) model for Sgr A*, where they
built a self-consistent jet-ADAF solution.  In this model, the
spectrum is mostly jet dominated, with comparable disk and jet
contributions in the X-ray and only a small optical/UV disk
component.  The same team's model for NGC4258 (Yuan et al. 2002)
is also jet-dominated, but requires a quite different approach
because of the need for a stronger shock to explain the more
powerful emission.  The low X-ray luminosity of IC~1459 compared
to the bright radio seems to be consistent with the idea that
this source is a low-luminosity AGN.

Second, this model suggests that some of the \.{M}$_{Bondi}$
could in fact be re-ejected in the form of a jet.  This would be
one explanation for the preponderance of strong jets in
low-luminosity systems, and is supported by theoretical models
and simulations.  For example, Livio, Ogilvie \& Pringle (1999)
have argued that the Blandford-Znajek mechanism may be enhanced
by additional poloidal field advection in ADAFs.  This is similar
to the conclusions of Meier (2001), who finds jet production
preferentially occurs in geometrically thick accretion flows
(i.e., ADAF/CDAF-like).  Thus, a weak disk may actually argue for
a stronger jet, as supported by the recent studies of both
stellar and galactic low-luminosity systems.

If there is accretion at the Bondi rate, and \.{M}$_{Bondi}$ is
processed through a standard accretion disk, then the maximum
disk temperature would be 25,600~K.  However, the disk fit of \S
\ref{markoff} is cool (7000~K). This implies a smaller
\.{M}=1.6$\times$10$^{-4}~M_{\odot}$~yr$^{-1}$ and hence an
unphysical $q_{jet}$=10. This rules out accretion through the
7000~K disk as the sole energy source for the jet, leaving
quasi-spherical accretion, or black hole spin energy (Blandford
\& Znajek 1977) as the remaining possibilities.  The low \.{M}
derived for the disk may be misleading, since the AGN-like blue
HST V-R color is consistent with a 25000~K disk. Better UV data are
needed to determine the maximum disk temperature, and so tie the disk
and jet physics together more closely.

The presence of H$\alpha$ implies the presence of Lyman continuum
photons, though the amount depends on the covering factor of the
H$\alpha$ producing gas to the continuum source, $f_c$. This UV
continuum could be provided either by an absorbed disk, or by the
jet itself. The combined inverse Compton component and post-shock
synchrotron component (which connects the X-ray to the high
frequency radio emission) imply a Lyman continuum flux of
10$^{-12}$~erg~cm$^{-2}$~s$^{-1}$.  
%
The agreement of the H$\alpha$/X-ray ratio in IC~1459 with those in
normal AGN is then surprising, since in those objects the Lyman
continuum is thought to come from a hot disk, while the X-rays
probably arise in a disk corona (Haardt \& Maraschi 1991, Nayakshin
2000). It may be that a jet component should be reconsidered in the
more normal AGN also.

\section{Conclusions}

We have presented \chandra\ observations of the nucleus of
IC~1459, a nearby (D=22~Mpc) elliptical galaxy with a measured
nuclear black hole mass of 2$\times$10$^9$M$_{\odot}$.  These
observations clearly separate out a weak
($L_X$=8$\times$10$^{40}$erg~s$^{-1}$, 0.3-8~keV), unabsorbed
nuclear X-ray source, with a slope $\Gamma$ = 1.88$\pm$0.09,
and no strong Fe-K line at 6.4~keV (EW$<$382~eV). This describes
a normal AGN X-ray spectrum, but lies at 3$\times$10$^{-7}$ below
the Eddington limit. The SED of the IC~1459 nucleus is extremely
radio loud compared to normal radio-loud quasars.

The nucleus is surrounded by hot ISM (kT$\sim$0.5-0.6~keV). We measure
an average density of this ISM of 0.31~cm$^{-3}$ within the central
1.7~arcsec radius (182~pc). The Bondi gravitational capture radius is
140~pc. (Although this could be resolved with \chandra, the nuclear
source prevents this measurement, so we do not have a direct measure
of the ISM density at the Bondi radius.) Using these gas parameters we
find the accretion rate at the gravitational capture radius, and
derive the luminosity if the accretion efficiency is at 10$\%$
L$_{acc}$$\sim$6$\times$10$^{44}$~erg~s$^{-1}$. Even allowing for
factors of a few uncertainties, L$_{acc}$ is orders of magnitude
higher than $L_X$.

We consider various possible explanations for this discrepancy:

(1) Obscuration. Scattering from a standard torus would produce
a strong Fe-K fluorescence line that is not seen. An alternative
4$\pi$ covering obscurer with small holes is ruled out by the
weakness of the nuclear infrared emission.

(2) Angular momentum impeded accretion. Sufficiently high
resolution hydrodynamic modelling of low angular momentum
accretion flows does not yet exist, however the indications are
that this would form a disk.  Since there are no predictions for
the resulting SED we cannot take this possibility further as yet.

(3) ADAF solutions. These can explain the X-ray spectrum, but
have trouble with the high radio/X-ray ratio in IC~1459, which is
much larger than in normal ellipticals, or even M~87. Quite
possibly an ADAF like solution leads to jet creation (ADIOS),
which we considered next.

(4) Jet models. We find that the jet model fits the
radio-100$\mu$m and X-ray spectra extremely well. The total power
in this jet is much larger than the radiated luminosity, and
amounts to $\sim$10\% of L$_{acc}$. Accretion close to the Bondi
rate is then needed.  
The model fitting uses a cool (7000~K) accretion disk, which does not
give a high enrough accretion rate to power the jet. The the disk may
be hotter, and better UV data could determine this. If this is not the
case the jet must be powered by quasi-spherical accretion, or
potentially by the black hole spin energy.
%


\acknowledgments

We thank the CXC DS and SDS teams for their efforts in reducing
the data and developing the software used for the reduction (SDP)
and analysis (CIAO).  We acknowledge enlightening discussions
with Fabrizio Fiore, and with Andrew King at the Aspen Center for
Physics summer workshop on `Compact Object Populations in
External Galaxies', and thank Gijs Verdoes-Kleijn for valuable
pointers to the HST data.  This research has made use of NASA's
Astrophysics Data System, and of the NASA/IPAC Extragalactic
Database (NED) which is operated by the Jet Propulsion
Laboratory, California Institute of Technology, under contract
with the National Aeronautics and Space Administration.  This
work was supported by NASA contract NAS~8--39073 (CXC),
\chandra\ GO grant NAS8-39073 and an NSF Astronomy and
Astrophysics Postdoctoral Fellowship (S.M.).



\clearpage

\begin{figure}
\plotone{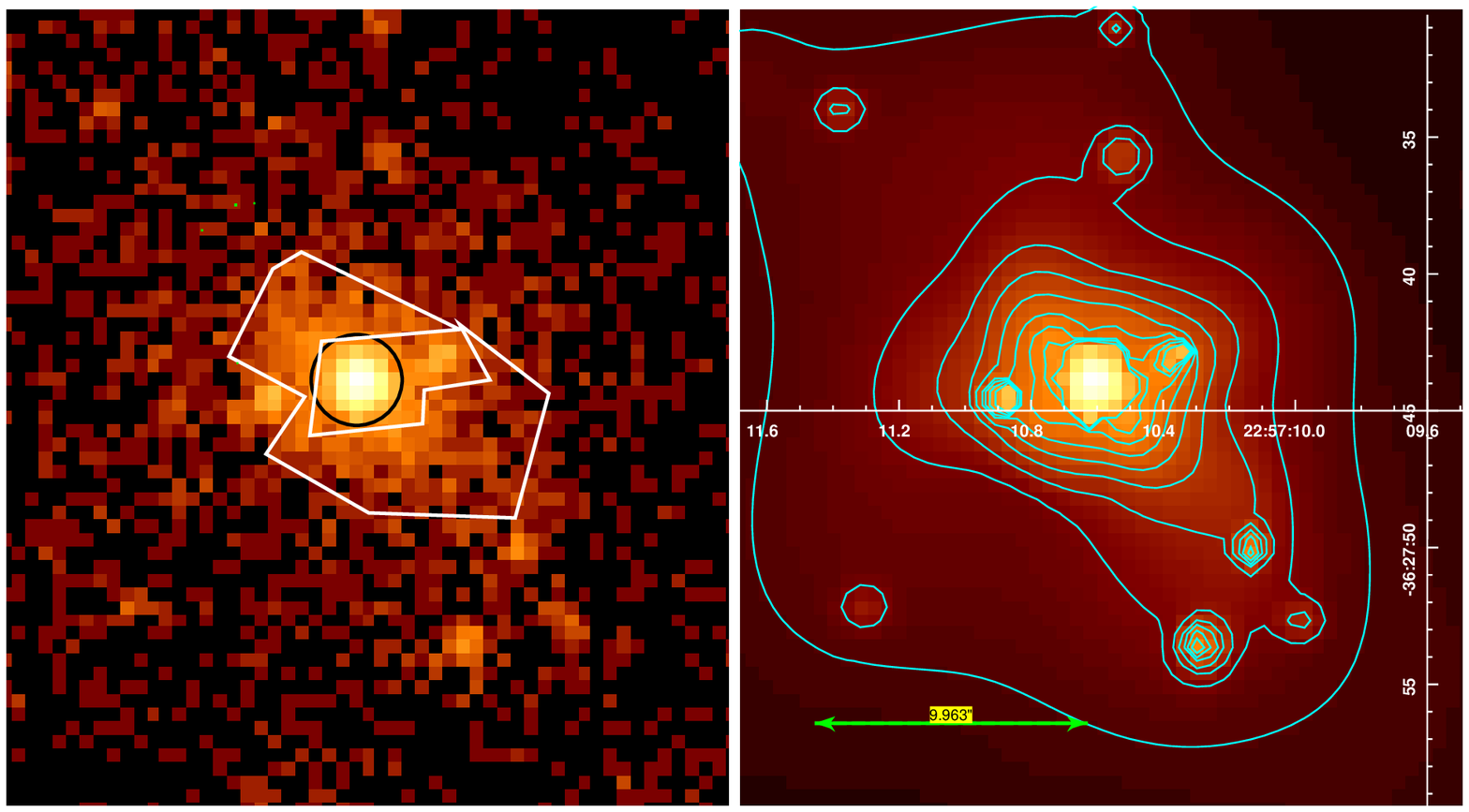}
\caption{ {\em Left:} {\em Chandra} X-ray (0.3-10~keV) ACIS-S3
image of the central area of IC~1459. The circle represent the
extraction region for the nuclear spectral counts. The polygon is
the extraction area for the circum-nuclear diffuse emission.
{\em Right:} Adaptively smoothed image of the same
region. Contours are logarithmically spaced from
0.011~counts/pixel to 20~counts/pixel. The horizontal bar is
10~arcsec long. \label{fig1}}
\end{figure}

\clearpage 

\begin{figure}
\plottwo{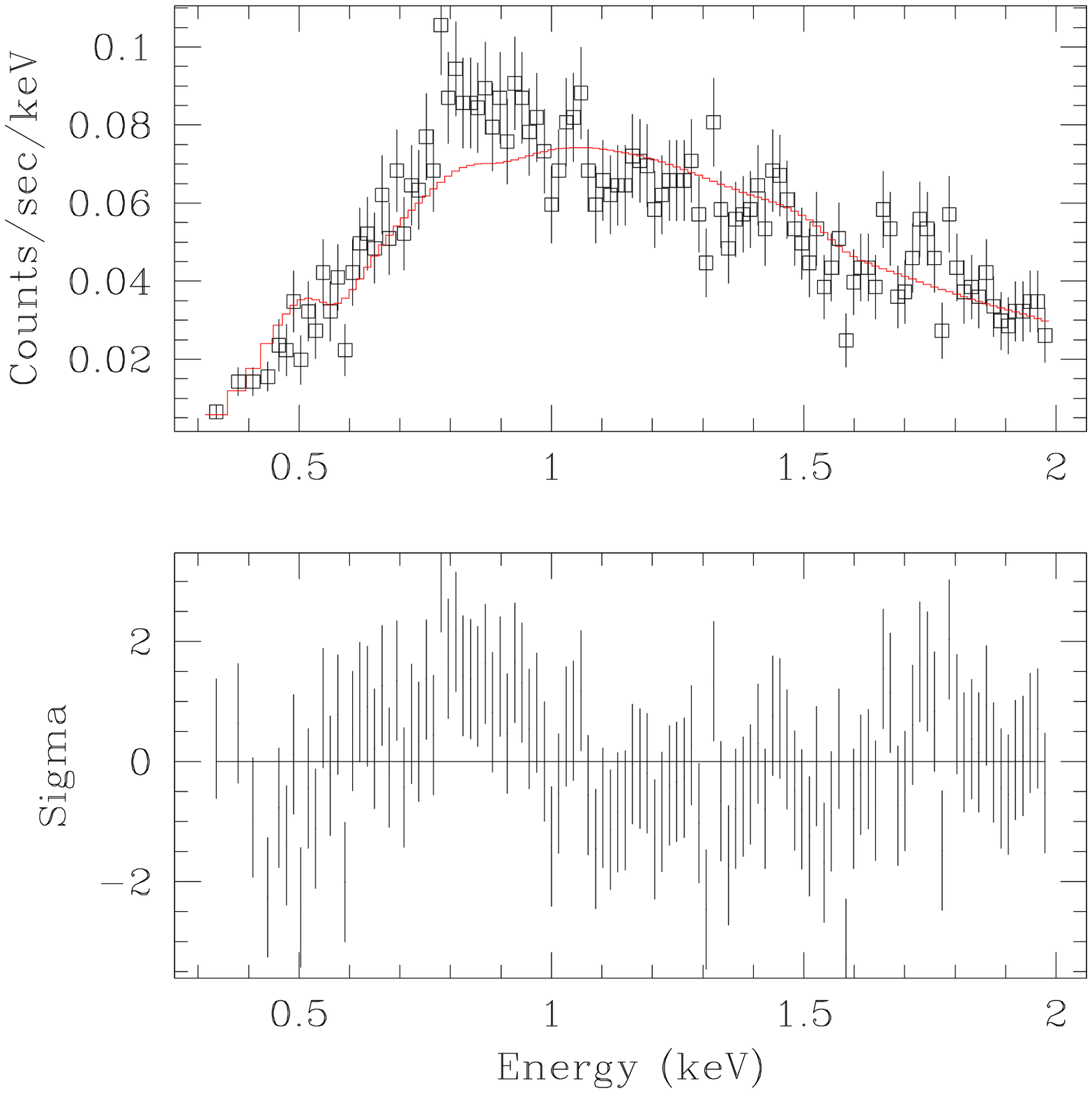}{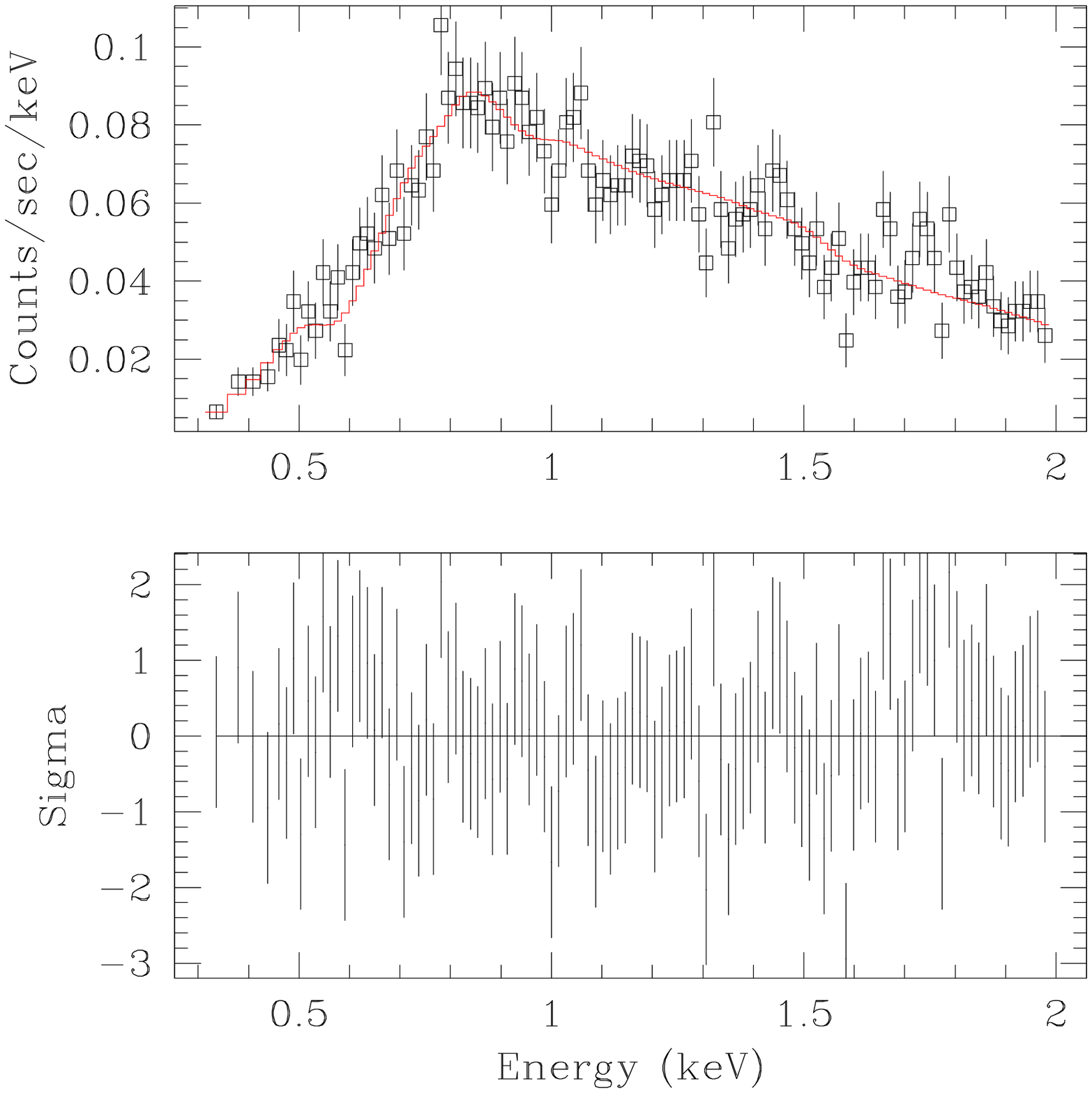}
\caption{
a) ACIS-S background-subtracted spectrum of the
nucleus of IC~1459 with best-fit single-power-law model (see
text), and fit residuals, plotted in the 0.3-2.0~keV energy
range.  b) same for the 2-component power-law + thermal emission
model. \label{fig2}}
\end{figure}

\clearpage

\begin{figure}
\plotone{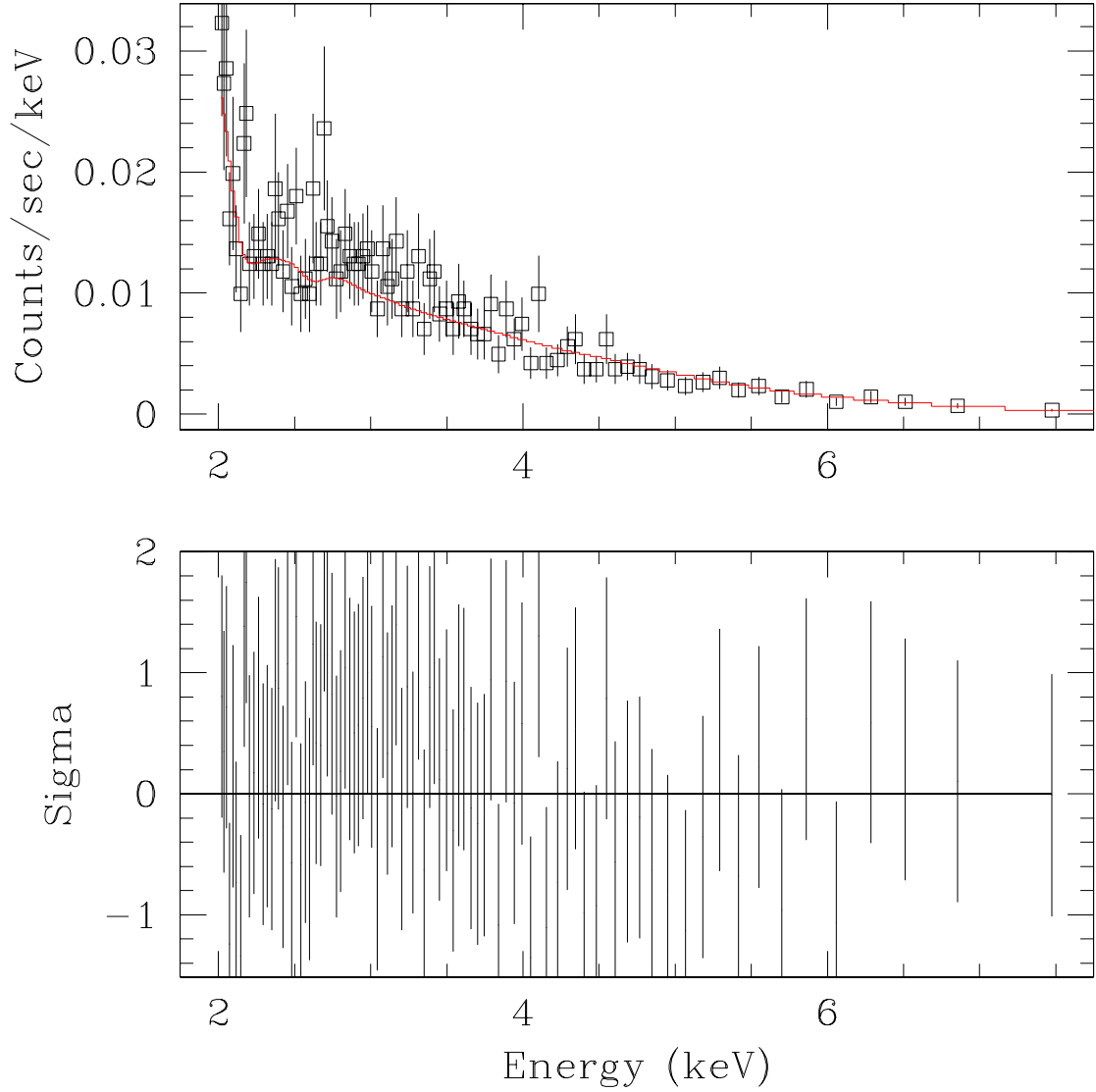}
\caption{ACIS-S background-subtracted spectrum of the nucleus
of IC~1459 with best-fit 2-component power-law + thermal emission
model and fit residuals, plotted in the 2.0-8.0~keV energy
range. \label{fig1}}
\end{figure}

\clearpage

\begin{figure}
\plotone{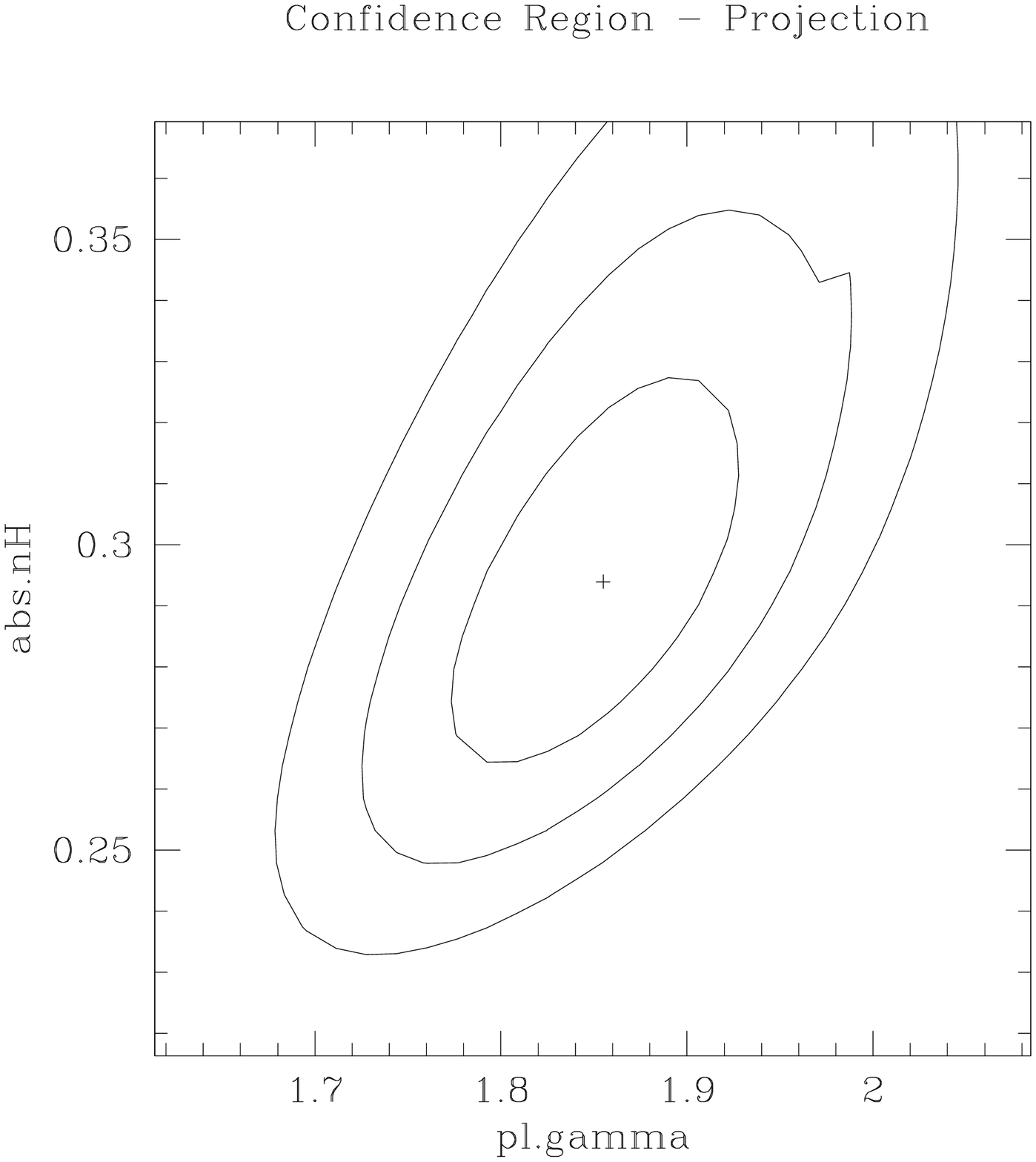}
\caption{$\Gamma / N_H$ confidence contours (1, 2, 3 $\sigma$
for 2 parameters) for the absorbed power-law component in the
power-law + thermal component fit. $N_H$ is in units of
10$^{22}$cm$^{-2}$. \label{fig1}}
\end{figure}

\clearpage

\begin{figure}
\plotone{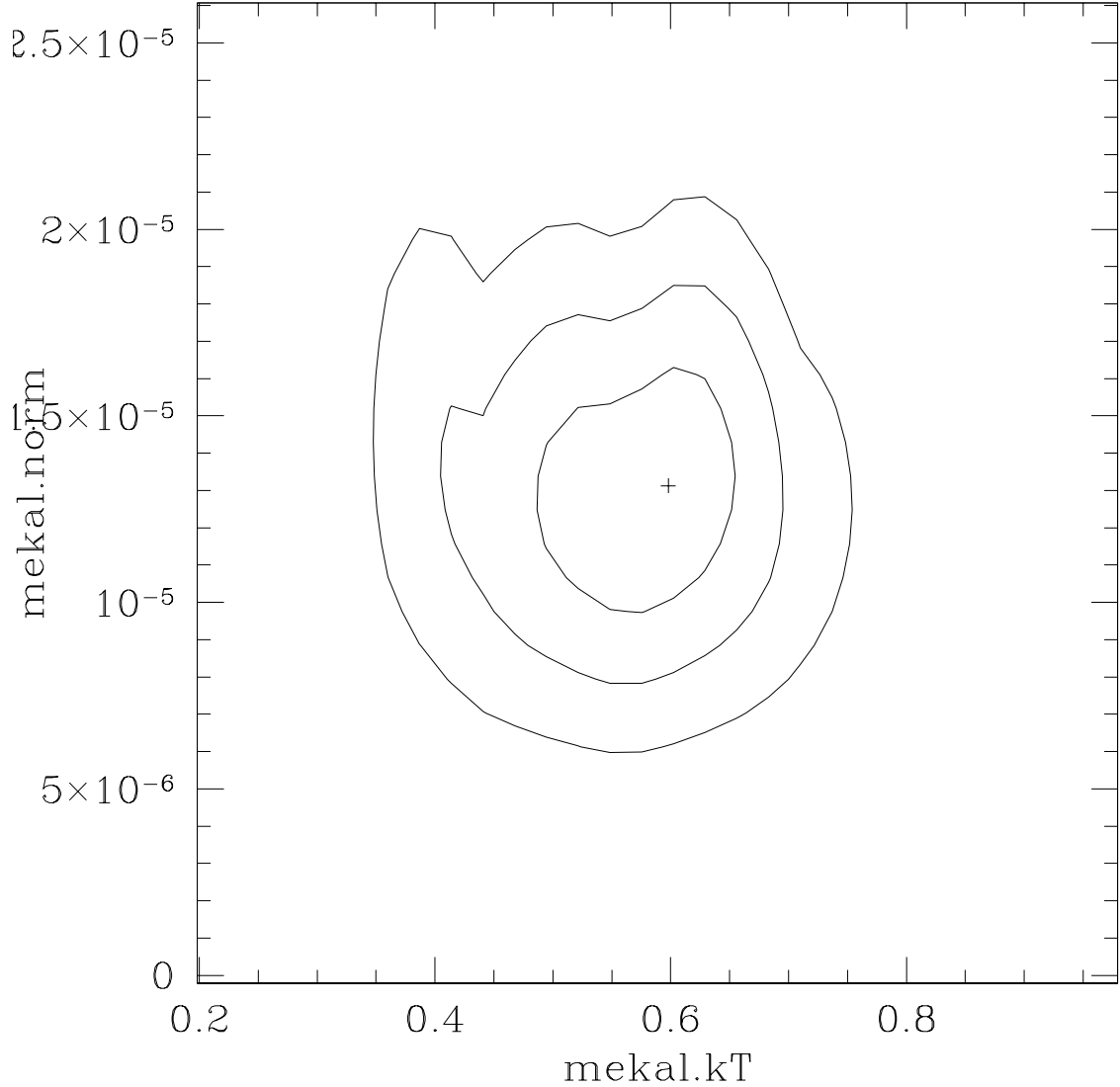}
\caption{kT / Normalization confidence contours (1, 2, 3
$\sigma$ for 2 parameters) for the {\tt mekal} thermal component
in the power-law + thermal component fit.  The temperature (kT)
of the thermal component is in keV. The normalization
(mekal.norm) is in units of $10^{-14} / 4 \pi D^2 $ * (emission
measure). The sharp edges are an artifact of the relatively small
number of points used to derive these contours.\label{fig1}}
\end{figure}

\clearpage

\begin{figure}
\plotone{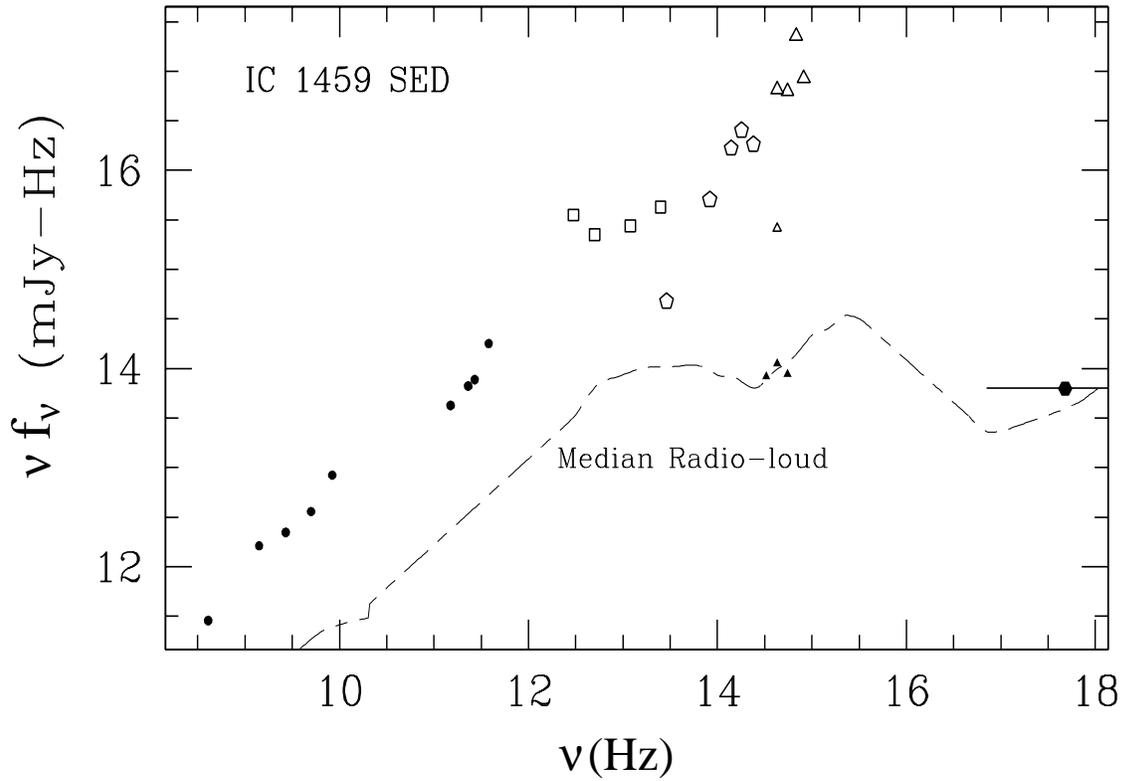}
\caption{Radio-to-X-rays Spectral Energy Distribution (SED) of
the nucleus of IC~1459: log of $\nu f_{\nu}$, where $\nu$ is the
frequency and $f_{\nu}$ is the flux at that frequency, is plotted
against log of the frequency $\nu$. In this representation $\nu
f_{\nu}$ gives the power emitted per logarithmic interval. The
dashed line is the median radio-loud quasar SED from Elvis et
al. (1994). (See text and table~4 for details of the data
points). \label{fig1}}
\end{figure}

\clearpage

\begin{figure}
\plotone{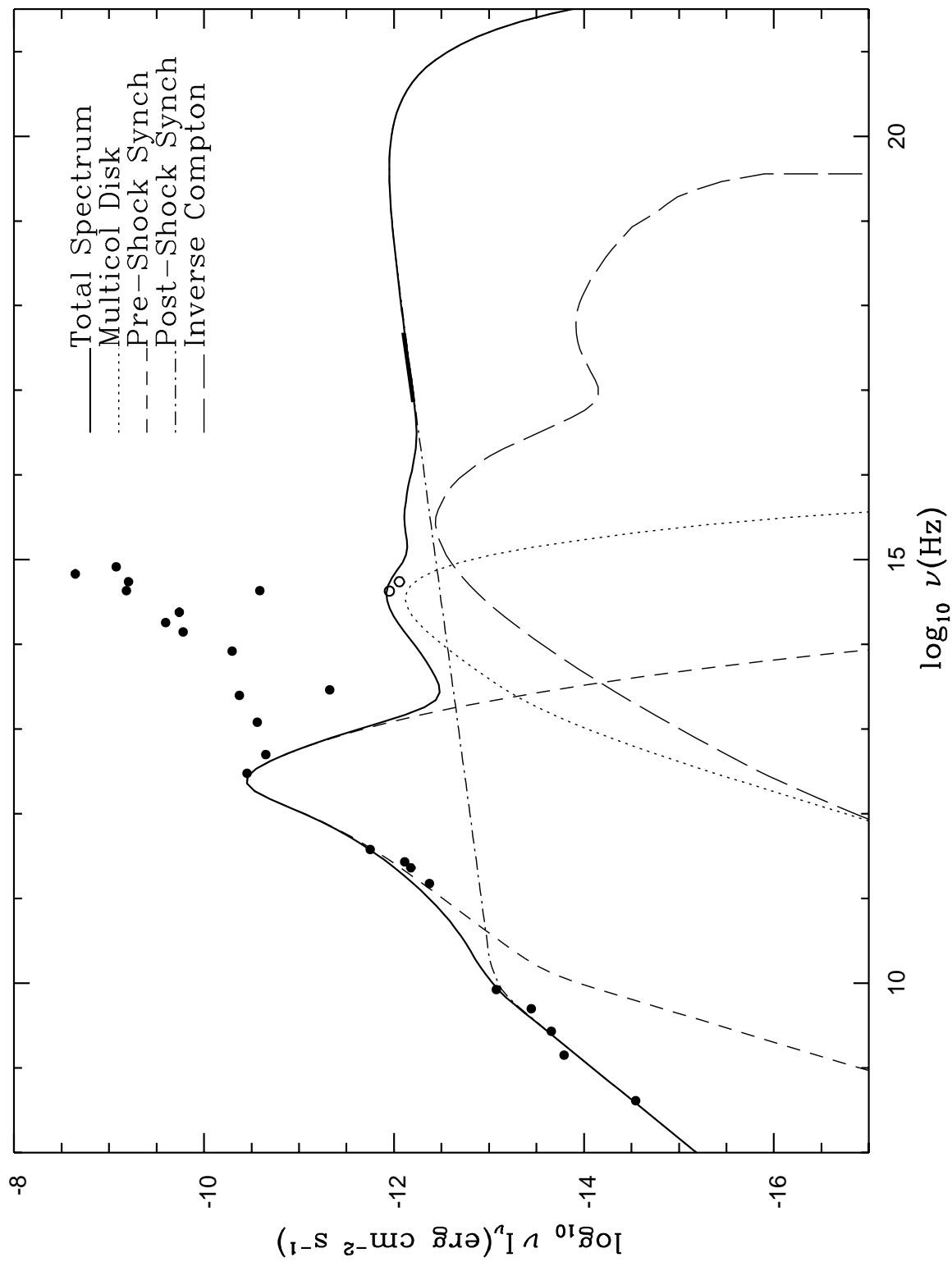}
\caption{Markoff-Falcke jet model fit to the SED of IC~1459.
(see text, \S\ref{markoff}, for details). \label{fig1}}
\end{figure}

\clearpage

\newpage
\centerline{Table 1}
\centerline{ IC~1459: Properties and Observation Log}

{\flushleft
\small
\begin{tabular}{ccccccc}
\hline 
Time& ObsID& D&Diam.&N$_H$&log$L_B$&log$M_{\bullet}^a$\\
(ks) &&(Mpc)&(')&(cm$^2$)&($L_{\odot}$)&($M_{\odot}$)\\
\hline
58.8& 2196&$ 22^b$&5&1.2E20&10.9&9.3\\
\hline
\end{tabular}

\noindent
a) log of nuclear massive black hole mass based on stellar
kinematics (Cappellari et al 2002, rescaled for D= 22~Mpc; this
is a factor of $\sim 6$ larger than the previous estimate based
on gas kinematics by Verdoes Kleijn et al 2000)

\noindent
b) Distance ($H_o = 75$) from Bender et al (1992). 
At this distance 1~arcsec correspnds to 107~pc.

\normalsize
\medskip
\centerline{Table 2}
\centerline{ IC~1459: Nuclear Spectral Parameters and Luminosity}

{\flushleft
\small
\begin{tabular}{cclccc}
\hline 
Model& Parameter$^a$& Best-Fit&$\chi^{2 b}$&$f_X^c/10^{-13}$&$L_X^c/10^{40}$\\
 &&($\pm 90\%$ errors)&dof&(ergs cm$^{-2}$ s$^{-1}$)&(ergs  s$^{-1})$\\
\hline
abs*pl& abs.nh (cm$^{-2}$)&$ 2.5 \times 10^{21}$&194.6&13.5&7.9\\
&&($\pm 0.20 \times 10^{21})$&195 & &\\
&$\Gamma$.pl&1.94&&&\\
&&($\pm 0.07$)&&&\\
&norm.pl$^d$$$&2.3$\times 10^4$&&&\\
&&($\pm 0.1 \times 10^4$)&&&\\
abs1*(mekal+abs*pl)$^e$&abs.nh (cm$^{-2}$)&$2.9 \times 10^{21}$&155.3&\\
&&(-0.3, +0.6)$\times 10^{21}$&192&&\\
&$\Gamma$.pl&1.88&&13.4&7.8\\
&&($\pm 0.09$)&&&\\
&norm.pl$^d$&$2.1 \times 10^-4$&&&\\
&&($\pm 0.2  \times 10^-4$)&&&\\
&kT.mekal (keV)&0.6&&0.34 &0.2\\
&&(-0.13, +0.10)&&&\\
&norm.mekal$^f$&$1.2 \times 10^{-5}$&&&\\
&&($\pm 0.3$)$\times 10^{-5}$&&&\\

\hline
\end{tabular}

\noindent
a) These are the fit parameters that were allowed to vary

\noindent
b) In the restricted (0.3-1.3)~keV range, $\chi^2 /dof$ = 78/59
for abs*pl and 38/57 for abs1*(mekal+abs*pl), where `dof' is the
number of degrees of freedom in the fit.

\noindent
c) In the (0.3-8)~keV range. Flux is at the source
unabsorbed. Luminosity is calculated for D=22~Mpc.

\noindent
d) Units are photons~keV$^{-1}$~cm$^{-2}$~s${-1}$ at 1~keV

\noindent
e) abs1.nh was set to $1.2 \times 10^{20} \rm cm^{-2}$, the Galactic
line of sight N$_H$.  The mekal parameter nh was frozen to the
default value of 1; the mekal abundance was assumed solar (frozen
to 1).

\noindent
f) norm.mekal = $10^{-14} / 4 \pi D^2 $ * (emission measure)

\normalsize

\medskip
\newpage

\centerline{Table 3} 
\centerline{ IC~1459: Spectral Parameters of the Diffuse Emission
}
\small
\begin{center}
\begin{tabular}{cclc}
\hline
Model& Parameter$^a$& Best-Fit&$\chi^{2}$\\
 &&($\pm 90\% errors$)&dof\\
\hline
abs*(mekal+pl)$^b$ &$\Gamma$.pl&1.6&50.7\\
&&($\pm 0.2$)&42\\
&norm.pl&$6.6 \times 10^{-6}$&\\
&&($\pm 1.2  \times 10^{-6}$)&\\
&kT.mekal (keV)&0.56&\\
&&(-0.07, +0.06)&\\
&norm.mekal$^c$&$8.2 \times 10^{-6}$&\\
&&($\pm 1.2$)$\times 10^{-6}$&\\
\hline
\end{tabular}
\end{center}

\noindent
a) These are the fit parameters that were allowed to vary

\noindent
b) abs1.nh was set to $1.2 \times 10^{20} \rm cm^{-2}$, the Galactic
line of sight N$_H$ (Stark et al. 1992).  See Table~2 for default
frozen mekal parameters.

\noindent
c) norm.mekal = $10^{-14} / 4 \pi D^2 $ * (emission measure)
\normalsize
\medskip

\newpage
\medskip
\centerline{Table 4} 
\centerline{SED for IC~1459}
\begin{center}
\begin{tabular}{llrrcr}
\hline
frequency(Hz) & band   &f(mJy)  &$\pm$ &  aperture ('')& ref \\
\hline
4.8e17        & 2~keV    &1.3e-4  & 0    &   2 & this paper \\
8.2e14        & U        & 103    & 0    &  31 & 6 \\
6.8e14        & B        & 333    & 0    &  31 & 6 \\
5.5e14        & V        & 114    & 0    &  31 & 6 \\
5.5e14        & V        & 0.16   & 0    & 0.1 & 7 \\
4.3e14        & R        & 0.26   & 0    & 0.1 & 8 \\
4.3e14        & R        & 153    & 0    &  31 & 6 \\
4.3e14        & R        &   6    & 0    &   1 & 5 \\
3.3e14        & I        & 0.25   & 0    & 0.1 & 7 \\
2.4e14        & J        &  76    & 0    &   5 & 1 \\
1.8e14        & H        & 141    & 0    &   5 & 1 \\
1.4e14        & K        & 119    & 0    &   5 & 1 \\
8.3e13        & L'       &  61    & 0    &   5 & 1 \\
2.9e13        & N        & 16.4   & 12.2 &   5 & 1 \\
2.5e13        & 12~$\mu$m& 170    & 29   &  60 & 3 \\
1.2e13        & 25~$\mu$m& 230    & 44   &  60 & 3 \\
5.0e12        & 60~$\mu$m& 450    & 31   &  60 & 3 \\
3.0e12        &100~$\mu$m& 1180   & 103  &  60 & 3 \\
3.8e11        & 0.8~mm   & 470    & 61   &  17 & 2 \\
2.7e11        & 1.1~mm   & 286    & 30   & 18.5& 2 \\
2.3e11        & 1.3~mm   & 289    & 19   & 19.5& 2 \\
1.5e11        & 2.0~mm   & 283    & 80   & 27.5& 2 \\
8.4e9         & 8.4~GHz  &1000    & 0    &   - & 4 \\
5.0e9         & 5.0~GHz  & 720    & 0    & 0.03& 9 \\
2.7e9         & 2.7~GHz  & 820    & 0    &   - & 4 \\
1.41e9        & 1.4~GHz  &1151    & 0    &   - & 4 \\
4.08e8        & 408~MHz  & 700    & 0    &   - & 4 \\
\hline
\end{tabular}
\end{center}
{\small references: 1. Sparks W.B. et al., 1986; 2. Knapp G.R. \&
Patten B.M., 1991 [JCMT]; 3. Jura M. et al. 1987 [IRAS];
4. PKSCAT90 [via NED], Wright A.E. \& Otrupek 1990 ({\tt
http://www.asc.rssi.ru/mdb/stars/8/8015.htm}); 5. Franx M.,
Illingworth G. \& Heckman T., 1998 [HST]; 6. Poulain P. 1988;
7. Carollo et al. 1997. [HST]; 8. Verdoes-Klein et
al. 2002. [HST]; 9. Slee et al. 1994. [PKS-Tidbinbilla].  }

\newpage
\medskip
\centerline{Table 5} 
\centerline{Model Fit Parameters}
\begin{center}
\begin{tabular}{lcr}
\hline
Parameter & Value \\
&\\
\hline
Power in jets& 2.8$\times$10$^{-4}~L_{Edd}$ \\
Accelerated electron energy index & 2.76 \\
Shock location along axis & 700 $r_g$ \\
Inclination angle & 30 degrees\\
Luminosity in blackbody disk &2.0$\times$10$^{-7}~L_{Edd}$ \\
T$_{in}$ for disk blackbody & 7.0$\times$10$^{3}$K \\

\hline
\end{tabular}

\end{center}

\end{document}